\documentclass[10pt,twocolumn,aps,pra,superscriptaddress,showpacs,amsmath,amssymb]{revtex4-1}

\usepackage{graphicx}
\usepackage{subfigure}
\usepackage[input-symbols={\pi\times}]{siunitx}
\usepackage{braket}
\usepackage{hyperref}

\allowdisplaybreaks 

\newcommand{\Tr}{\mathrm{Tr}}

\begin{document}

\title{State-selective all-optical detection of Rydberg atoms}

\author{Florian Karlewski} 
\email[]{florian.karlewski@uni-tuebingen.de}
\affiliation{CQ Center for Collective Quantum Phenomena and their Applications, Physikalisches Institut, Eberhard-Karls-Universit\"at T\"ubingen, Auf der Morgenstelle 14, D-72076 T\"ubingen, Germany} 

\author{Markus Mack} 
\affiliation{CQ Center for Collective Quantum Phenomena and their Applications, Physikalisches Institut, Eberhard-Karls-Universit\"at T\"ubingen, Auf der Morgenstelle 14, D-72076 T\"ubingen, Germany} 

\author{Jens Grimmel}
\affiliation{CQ Center for Collective Quantum Phenomena and their Applications, Physikalisches Institut, Eberhard-Karls-Universit\"at T\"ubingen, Auf der Morgenstelle 14, D-72076 T\"ubingen, Germany} 

\author{N\'{o}ra S\'{a}ndor}
\altaffiliation{Present affiliation: IPCMS (UMR 7504) and ISIS (UMR 7006), Universit´e de Strasbourg and CNRS, Strasbourg, France}
\affiliation{CQ Center for Collective Quantum Phenomena and their Applications, Physikalisches Institut, Eberhard-Karls-Universit\"at T\"ubingen, Auf der Morgenstelle 14, D-72076 T\"ubingen, Germany} 
\affiliation{Institute for Solid State Physics and Optics, Wigner Research Centre for Physics, Hungarian Academy of Sciences, H-1525 Budapest P.O. Box 49, Hungary}

\author{J\'{o}zsef Fort\'{a}gh}
\email[]{fortagh@uni-tuebingen.de}
\affiliation{CQ Center for Collective Quantum Phenomena and their Applications, Physikalisches Institut, Eberhard-Karls-Universit\"at T\"ubingen, Auf der Morgenstelle 14, D-72076 T\"ubingen, Germany} 

\date{\today}
\begin{abstract}
	We present an all-optical protocol for detecting population in a selected Rydberg state of alkali atoms. 
The detection scheme is based on the interaction of an ensemble of ultracold atoms with two laser pulses: one weak probe pulse which is resonant with the transition between the ground state and the first excited state, and a pulse with high intensity which couples the first excited state to the selected Rydberg state. We show that by monitoring the absorption signal of the probe laser over time, one can deduce the initial population of the Rydberg state. Furthermore, it is shown that -- for suitable experimental conditions -- the dynamical absorption curve contains information on the initial coherence between the ground state and the selected Rydberg state. We present the results of a proof-of-principle measurement performed on a cold gas of $^{87}$Rb atoms. The method is expected to find application in quantum computing protocols based on Rydberg atoms.
\end{abstract}

\maketitle

\section{Introduction\label{sec:intro}}

Rydberg atoms coupled to electromagnetic fields form a promising system for the physical realization of quantum information protocols~\cite{Saffman.2010} and quantum simulations~\cite{Weimer.2010}. In these protocols qubits are realized by a set of atomic states, which includes one or potentially more Rydberg levels. One requirement of these schemes is the ability to measure the Rydberg states' population in order to read out the results of the quantum operations. For accomplishing this task, most experiments with ultracold Rydberg gases use methods including field ionization and subsequent detection of electrons and/or ions on multi-channel plates or channeltrons~\cite{Low.2012}. These techniques offer high sensitivity and -- for carefully chosen experimental conditions~\cite{Tate.2007,Caliri.2007} -- selectivity among the Rydberg levels~\cite{Oliveira.2002,Nascimento.2006,Branden.2010}. 

Selective field ionization (SFI) techniques are based on the fact that the ionization threshold is different for each atomic state, increasing from higher to lower lying levels. Hence, by slowly ramping up the electric field and monitoring the electrons/ions over time it is possible to deduce the initial populations in each level. However, the population of a lower lying Rydberg level cannot be probed without destroying the population of any higher lying Rydberg state. Therefore this method is not applicable in protocols which require independent probing of multiple Rydberg states\cite{Cano.2014}. Another inherent property of methods based on ionization is that the detected atoms are removed from the system and cannot be reused. Although this atomic loss is negligible in most cases \cite{Low.2012}, it might be a serious limitation in experiments working with only one or a few atoms \cite{Labuhn.2014}. 

One alternative to ionization detection methods is all-optical probing based on electromagnetically induced transparency (EIT)~\cite{Fleischhauer.2005}. This approach has been successfully applied in order to non-destructively probe the Rydberg level structure in non-interacting~\cite{Mohapatra.2007, Mack.2011} and weakly interacting~\cite{Weatherill.2008} Rydberg gases, as well as in the presence of electric fields~\cite{Tauschinsky.2010,Bason.2010, Hattermann.2012, Tauschinsky.2013,Grimmel.2015}. These experiments, however, did not access the population of the Rydberg state. On the other hand, an EIT-based scheme for the optical detection of Ryd\-berg population~\cite{Gunter.2012,Gunter.2013} has been proposed and demonstrated in dense atomic clouds where the Rydberg blockade allows the spatially resolved detection of Rydberg atoms. 

Here we propose an all-optical scheme for detecting the population in a selected Rydberg state in dilute gases. By using a series of laser pulses in EIT configuration this technique also allows for distinction between coherent superpositions and statistical mixtures of the ground and Rydberg states of the atoms. Since this scheme is based on time-resolved observation of the optical response of individual atoms it may, in principle, be used down to the single atom level. Our method is state selective and applicable for testing the population not only in the highest Rydberg level of interest (cf. SFI) but any lower lying or intermediate Rydberg state. 

We present our theoretical model along with numerical simulations and demonstrate the scheme in a proof-of-principle experiment with a dilute gas of $^{87}$Rb atoms showing the detection of the population in an initially prepared Rydberg state. Our analysis includes characteristic effects of Rydberg experiments such as blackbody-induced depopulation \citep{Beterov.2009,Beterov.2009.Err}, superradiance \cite{Wang.2007} and dipole-dipole interaction \cite{Anderson.2002}.

\section{Theoretical model\label{sec:mathmodel}}

\begin{figure}
	\includegraphics{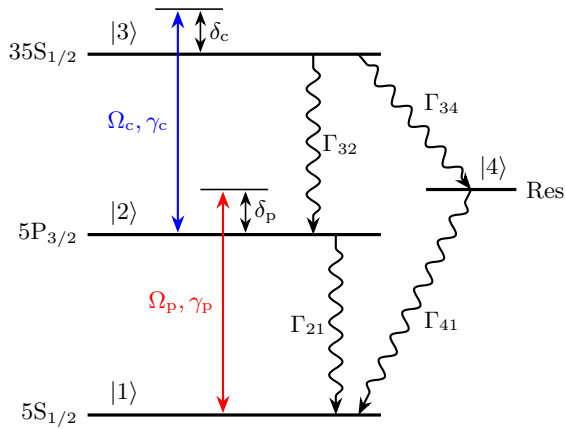}
	\caption{\label{fig:ladder}(Color online) Illustration of the atom-laser interaction as used for the model of time-resolved electromagnetically induced transparency (EIT). States $\ket{1}$, $\ket{2}$ and $\ket{3}$ denote the ground state, the first excited state and a Rydberg state of the atom, respectively. In our experimental setup, these states correspond to the $5\mathrm{S}_{1/2} (F=2)$, $5\mathrm{P}_{3/2} (F=3)$ and $35\mathrm{S}_{1/2} (F=2)$ states of $^{87}$Rb. The atomic transition $\ket{1} \leftrightarrow \ket{2}$ is driven by a weak probe laser with the Rabi frequency $\Omega_\mathrm{p}$ (red), while the transition $\ket{2} \leftrightarrow \ket{3}$ is driven by a stronger coupling laser with the Rabi frequency $\Omega_\mathrm{c}$ (blue). $\gamma_\mathrm{p}$ and $\gamma_\mathrm{c}$ denote coherence decay terms. $\delta_\mathrm{p}$ and $\delta_\mathrm{c}$ are the detunings of each laser to the corresponding atomic resonance. The radiative decay from the selected Rydberg state $\ket{3}$ to the neighboring states is accounted for by including a reservoir state $\ket{4}$. $\Gamma_{ij}$ denote the respective incoherent decays, consisting of spontaneous emission as well as transitions induced by blackbody radiation.}
\end{figure}

Let us consider a cold atomic gas interacting with two laser pulses in an EIT-like configuration~\cite{Fleischhauer.2005}. One of the pulses, resonant with the atomic transition between the ground state $\ket{1}$ and the first excited state $\ket{2}$, is a weak probe pulse well below the saturation intensity, while the other one is a relatively strong coupling pulse which is resonant with the atomic transition between the first excited state $\ket{2}$ and a selected (arbitrary) Rydberg $n\mathrm{S}_{1/2}$ state $\ket{3}$ (see Fig.~\ref{fig:ladder}). A single ``reservoir'' state $\ket{4}$ is used to model the neighboring Rydberg states~\cite{Weatherill.2008}. 

We use a semiclassical approach for describing the dynamics of the system and the laser pulses are taken into account through their classical electric field. The atomic gas is modeled by a motionless ensemble of atoms. The state of the atoms is described by the density matrix $\rho=\sum_{i,j=1}^4{\rho_{i,j}\ket{i}\bra{j}}$, where the states $\ket{2}$ and $\ket{3}$ rotate with the atomic transition frequencies $\omega_{21}$ and $\omega_{23}$, respectively. 

The time evolution of $\rho(t)$ is described by the master equation
\begin{equation}
	i\hbar\dot{\rho}=\left[\mathcal{H},\rho\right]+\mathcal{U}[\rho].
	\label{eqn:master} 
\end{equation}
Here the Hamiltonian $\mathcal{H}$ accounts for the interactions between the atoms and the laser pulses. The effects due to interatomic interactions are considered through dynamic effective rates in $\mathcal{U}[\rho]$ along with radiative losses occurring in the system. The Hamiltonian $\mathcal{H}$ is written as 
\begin{align}
	\mathcal{H}=&-\frac{\hbar}{2}\left(\Omega_\mathrm{p}\ket{2}\bra{1}+\Omega_\mathrm{c}\ket{3}\bra{2}+\mathrm{h.c.}\right)\nonumber\\
	&-\hbar\left( \delta_\mathrm{p}\ket{2}\bra{2}+\delta_\mathrm{c}\ket{3}\bra{3}\right),
	\label{eqn:Hamiltonian}
\end{align}
where $\Omega_\mathrm{p}=(E_\mathrm{p}d_{12})/\hbar$ and $\Omega_\mathrm{c}=(E_\mathrm{c}d_{23})/\hbar$ are the Rabi frequencies of the probe and coupling lasers, with $E_\mathrm{p}$ and $E_\mathrm{c}$ being the electric fields, and $d_{12}$ and $d_{23}$ the dipole matrix elements of the corresponding transitions, whereas $\delta_\mathrm{p}$ and $\delta_\mathrm{c}$ are the detunings of the probe and coupling laser from the corresponding transitions, respectively (see Fig.~\ref{fig:ladder}). Although we consider a situation where both the coupling and the probe laser are resonant with the atomic transitions they drive (i.e. $\delta_\mathrm{p}=\delta_\mathrm{c}=0$), by including these detunings one can account for potentially uncompensated electric and/or magnetic fields in a specific experimental realization. The operator $\mathcal{U}$ which governs the non-Hamiltonian part of the dynamics reads as
\begin{align}
	\mathcal{U}[\rho]=\quad&\frac{\Gamma_{32}}{2}\left(2\sigma_{13}\rho\sigma_{31}-\sigma_{33}\rho-\rho\sigma_{33}\right)\nonumber\\
	+&\frac{\Gamma_{21}}{2}\left(2\sigma_{12}\rho\sigma_{21}-\sigma_{22}\rho-\rho\sigma_{22}\right)\nonumber\\
	+&\frac{\Gamma_{34}}{2}\left(2\sigma_{43}\rho\sigma_{34}-\sigma_{33}\rho-\rho\sigma_{33}\right)\nonumber\\
	+&\frac{\Gamma_{41}}{2}\left(2\sigma_{14}\rho\sigma_{41}-\sigma_{44}\rho-\rho\sigma_{44}\right)\nonumber\\
	+&\frac{\gamma_\mathrm{p}}{2}\left(2\sigma_{11}\rho\sigma_{11}-\rho_{11}\sigma-\rho\sigma_{11}\right)\nonumber\\
	+&\frac{\gamma_\mathrm{c}}{2}\left(2\sigma_{33}\rho\sigma_{33}-\rho_{33}\sigma-\rho\sigma_{33}\right),
	\label{eqn:Lindblad}
\end{align}
where $\sigma_{kj}=\ket{k}\bra{j}$ are the atomic projection operators ($k,j\in\{1,2,3,4\}$). 

There are multiple sources of non-unitary dynamics in the system. One of them is the spontaneous emission from the first excited state $\ket{2}$ and the Rydberg state $\ket{3}$ which we take into account by introducing radiative decay rates $\Gamma_{21}$ and $\Gamma_{32}$. Another source, if present, is a depopulation of the Rydberg state $\ket{3}$ towards the neighboring Rydberg states. The depopulation may occur due to several phenomena depending on the actual realization of the system, such as amplified spontaneous emission and/or superradiance~\cite{Wang.2007, Day.2008} as well as induced emission and absorption due to the blackbody radiation of the environment~\cite{Beterov.2009,Beterov.2009.Err}. Following \cite{Day.2008}, we take these effects into account by modifying the third term in Eq.~\eqref{eqn:Lindblad} to the dynamic effective decay rate 
\begin{equation}
	\tilde{\Gamma}_{34}(t) = \Gamma_{34,\mathrm{sp}} \cdot \left[\rho_{44}(t)\cdot p_\mathrm{sup} + 1\right] + \Gamma_{34,\mathrm{bb}},\label{eqn:superradiance}
\end{equation}
with $p_\mathrm{sup}$ a superradiance parameter, and $\Gamma_{34,\mathrm{sp}}$ and $\Gamma_{34,\mathrm{bb}}$ the effective decay rates caused by spontaneous emission and blackbody radiation, respectively. In our model we assume that the entire population eventually ends up in the ground state $\ket{1}$. This assumption is valid provided the ionization from all involved states is negligible.

The above mentioned phenomena cause population-transfer between the atomic states. In contrast, there are a group of processes which do not result in a significant energy-decay in the system but leads to a relevant coherence loss. One such process is the phase noise of the driving lasers, which is included into the model through the coherence decay rates $\gamma_\mathrm{p}$ and $\gamma_\mathrm{c}$. Due to the redistribution of population from $\ket{3}$ to $\ket{4}$, atoms in Rydberg $n\mathrm{P}_j$ states are present in the cloud at various distances. As observed by~\cite{Anderson.2002}, the dipole-dipole interaction with these $n\mathrm{P}_j$ state atoms results in an inhomogeneous broadening of the Rydberg $n\mathrm{S}_{1/2}$ state. We take this into account by adding an effective dephasing term in Eq.~\eqref{eqn:Lindblad} to $\gamma_\mathrm{c}$, 
\begin{equation}
	\tilde{\gamma}_\mathrm{c}(t) = \gamma_\mathrm{c} + \gamma_\mathrm{3,dd} \cdot \rho_{44}(t). \label{eqn:pnoise}
\end{equation} 

The optical response of the cloud under the effect of the two laser pulses is given by the macroscopic polarization $\vec{P}=\mathcal{N}\Tr[\rho\vec{d}]$ where $\vec{d}=\sum_{i\neq j}\left(d_{ij}\ket{i}\bra{j}+\mathrm{h.c.}\right)$ is the atomic dipole operator. The absorption $\alpha$ of the probe laser is then given by the imaginary part of the electric susceptibility $\chi$, 
\begin{equation}
	\alpha(t)=\Im({\chi(t)})=\frac{\mathcal{N}d_{12}^2}{\epsilon_0\hbar\Omega_\mathrm{p}}\Im(\rho_{21}(t)),\label{eqn:abs}
\end{equation}
where $\mathcal{N}$ is the atom density of the cloud. Note that here we make the approximation that the cloud is homogeneously irradiated and the propagation effects of the laser pulses can be neglected. The time dependent absorption signal is thus given by the master equation, which, using the operators given in Eqs.~\eqref{eqn:Hamiltonian} and~\eqref{eqn:Lindblad}, reads as
\begin{subequations}
	\label{eqn:mastereqsys}
	\begin{align}
		\dot\rho_{11}=&\frac{i}{2}\left(\Omega_\mathrm{p}^{\ast}\rho_{21}-\Omega_\mathrm{p}\rho_{12}\right)+\Gamma_{21}\rho_{22}+\Gamma_{41}\rho_{44}\\
		\dot\rho_{22}=&\frac{i}{2}\left(\Omega_\mathrm{p}\rho_{12}-\Omega_\mathrm{p}^{\ast}\rho_{21}-\Omega_\mathrm{c}\rho_{23}+\Omega_\mathrm{c}^{\ast}\rho_{32}\right)\nonumber\\
		&-\Gamma_{21}\rho_{22}+\Gamma_{32}\rho_{33}\\
		\dot\rho_{33}=&\frac{i}{2}\left(\Omega_\mathrm{c}\rho_{23}-\Omega_\mathrm{c}^{\ast}\rho_{32}\right)-(\Gamma_{32}+\tilde{\Gamma}_{34})\rho_{33}\\
		\dot\rho_{21}=&\frac{i}{2}\left[\Omega_\mathrm{c}^{\ast}\rho_{31}-\Omega_\mathrm{p}\left(\rho_{22}-\rho_{11}\right)+2\delta_\mathrm{p}\rho_{21}\right]\nonumber\\
		&-\frac{1}{2}\left(\Gamma_{21}+\gamma_\mathrm{p}\right)\rho_{21}\\
		\dot\rho_{31}=&\frac{i}{2}\left[\Omega_\mathrm{c}\rho_{21}-\Omega_\mathrm{p}\rho_{32}+2(\delta_\mathrm{p}+\delta_\mathrm{c})\rho_{31}\right]\nonumber\\
		&-\frac{1}{2}(\gamma_\mathrm{p}+\tilde{\gamma}_\mathrm{c}+\Gamma_{32}+\tilde{\Gamma}_{34})\rho_{31}\\
	    \dot\rho_{32}=&\frac{i}{2}\left[-\Omega_\mathrm{p}^{\ast}\rho_{31}-\Omega_\mathrm{c}\left(\rho_{33}-\rho_{22}\right)-2\delta_\mathrm{c}\rho_{32}\right]\nonumber\\
		&-\frac{1}{2}\left(\tilde{\gamma}_\mathrm{c}+\Gamma_{21}+\Gamma_{32}\right)\rho_{32}\\
		\dot\rho_{44}=&\tilde{\Gamma}_{34}\rho_{33}-\Gamma_{41}\rho_{44}\\
		\dot\rho_{41}=&\frac{i}{2}\left(-\Omega_\mathrm{p}\rho_{42}+2\delta_\mathrm{p}\rho_{41}\right)-\frac{1}{2}(\Gamma_{41}+\gamma_\mathrm{p})\rho_{41}\\
		\dot\rho_{42}=&-\frac{i}{2}(\Omega_\mathrm{p}^{\ast}\rho_{41}+\Omega_\mathrm{c}\rho_{43})-\frac{1}{2}\left(\Gamma_{21}+\Gamma_{41}\right)\rho_{42}\\
		\dot\rho_{43}=&-\frac{i}{2}(\Omega_\mathrm{c}^{\ast}\rho_{42}+2\delta_\mathrm{c}\rho_{43})\nonumber\\
		&-\frac{1}{2}\left(\Gamma_{32}+\tilde{\Gamma}_{34}+\Gamma_{41}+\tilde{\gamma}_\mathrm{c}\right).
	\end{align}
\end{subequations}
When there is no coupling to the Rydberg state ($\Omega_\mathrm{c}=0$), an analytical steady state solution for $\alpha$ can be obtained, which will be used for normalization: 
\begin{align}
	\alpha_0=&\frac{\mathcal{N}d_{12}^2}{\epsilon_0\hbar}\cdot\Gamma_{21}(\gamma_\mathrm{p}+\Gamma_{21})\nonumber\\
	&\times\left[\gamma_\mathrm{p}^2\Gamma_{21}+2\gamma_\mathrm{p}\left(\Omega_\mathrm{p}^2+\Gamma_{21}^2\right)\right.\nonumber\\
	&\quad\left.+\Gamma_{21}\left(4\delta_\mathrm{p}^2+2\Omega_\mathrm{p}^2+\Gamma_{21}^2\right)\right]^{-1}.\label{eqn:alpha0}
\end{align}

\section{Detection of the Rydberg population: results of the numerical simulation\label{sec:simres}}

\begin{figure}
	\includegraphics{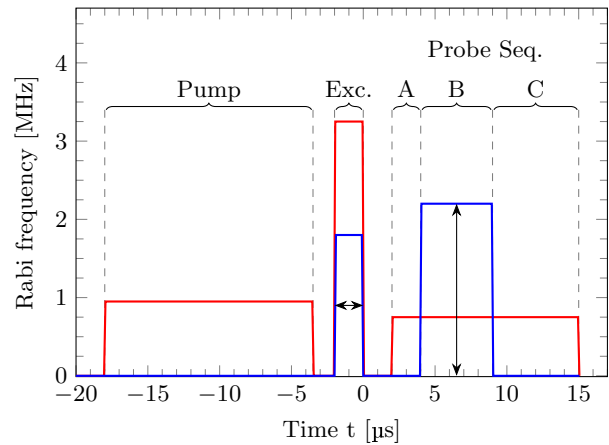}
	\caption{\label{fig:PulseSequence}(Color online) Pulse sequence for the coupling (blue) and probe (red) lasers. At $t=0$ a fraction of the atomic population is prepared in the Rydberg state. In the simulations this fraction is an input value, while in the experiment it is determined by the length of the excitation pulse (Exc.). Before this, an optical pumping pulse (Pump) is used to pump the atoms to the correct polarization. The time evolution of the optical density is monitored after the excitation pulse with the probe laser (A, B, C). The coupling laser is added in time interval B (EIT pulse). The Rabi frequencies are taken from the experimental values in part~\ref{sec:exp_setup}.}
\end{figure}

\begin{figure}
	\includegraphics{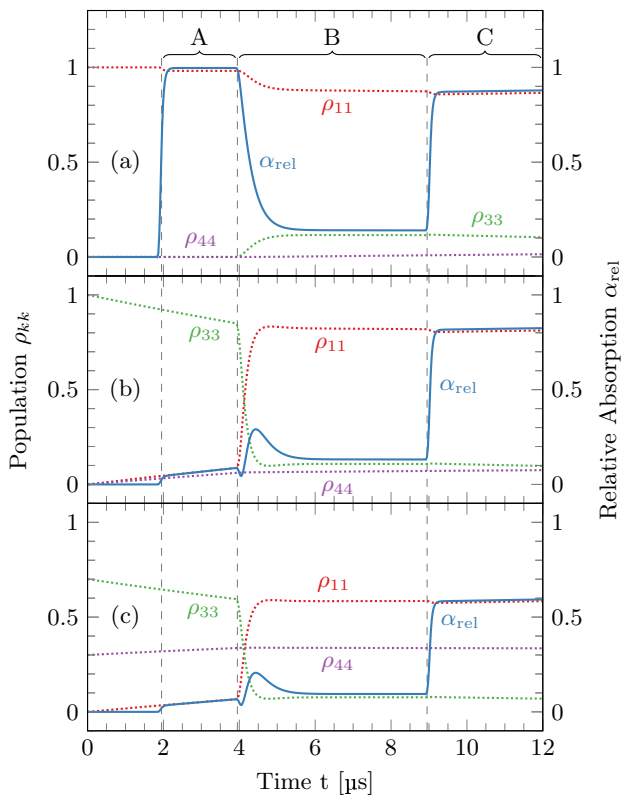}
	\caption{\label{fig:popalpha}(Color online) Dynamics of the populations $\rho_{kk}(t)$ of the atomic states (dotted lines) and the relative absorption $\alpha_\mathrm{rel}(t)$ (solid lines) of the probe laser induced by the pulse sequence A, B, C (see Fig.~\ref{fig:PulseSequence}) for atoms initially prepared in the ground state (a), the selected Rydberg state (b) and the reservoir state (c). Since the population $\rho_{22}$ is almost zero at all times, except for a transient population in the first \SI{100}{\nano\second} of B, it is not shown here.} 
\end{figure}

We numerically solve the equation system~\eqref{eqn:mastereqsys} with the pulse sequence of the lasers given by $\Omega_\mathrm{p}(t)$ and $\Omega_\mathrm{c}(t)$ (see the sequence A, B, C in Fig.~\ref{fig:PulseSequence}). The solution provides a description of the population dynamics while the atom is being probed by the weak laser on the $\ket{1}\leftrightarrow\ket{2}$ transition along with a time-dependent coupling between states $\ket{2}$ and $\ket{3}$. Furthermore, through Eq.~\eqref{eqn:abs} it describes the absorption of the probe laser, which we give relative to $\alpha_0$ (see Eq.~\eqref{eqn:alpha0}): 
\begin{equation}
	\alpha_\mathrm{rel}(t)=\frac{\alpha(t)}{\alpha_0}
\end{equation}
In Fig.~\ref{fig:popalpha} we show results for the cases where the entire population is initially (a) in the ground state ($\rho_0\equiv\rho(t=0)=\ket{1}\bra{1}$) or (b) in the selected Rydberg state ($\rho_0=\ket{3}\bra{3}$). In (c) the population is split between the Rydberg state and the reservoir state ($\rho_0=0.7\ket{3}\bra{3}+0.3\ket{4}\bra{4}$). 

Following the pulse sequence, the time evolution of the system can be separated into three major parts. When the atoms are initially prepared in the ground state, only small changes in the populations are visible (see Fig. \ref{fig:popalpha}(a)). However, the reasons for these changes are not the same in the different parts of the time evolution. In part A, when the atoms are only irradiated by the relatively weak probe laser ($\Omega_\mathrm{p}\ll\Gamma_{21}$), a small fraction of the population is transferred to the first excited state $\ket{2}$ by the absorbed light. In part B, the population transfer to state $\ket{2}$ is prevented by the strong coupling laser applied on the transition between states $\ket{2}$ and $\ket{3}$ and the absorption is reduced, which is the well-known effect of EIT. The timescale for the transparency to build up is defined by the Rabi-frequency $\Omega_\mathrm{c}$ of the coupling laser. If the requirement $\Omega_\mathrm{p}\ll\Omega_\mathrm{c}$ is not fulfilled, the transparency is only partial. In this case, the two laser fields cause two-photon transitions to state $\ket{3}$, and the absorption of the probe pulse is nonzero. This absorption level (in case of $\delta_\mathrm{p}=\delta_\mathrm{c}=0$) depends on $\Omega_\mathrm{p}/\Omega_\mathrm{c}$ and the decoherences $\gamma_\mathrm{p}$ and $\gamma_\mathrm{c}$. A consequence of this effect is that different initial populations of Rydberg states cause a different absorption level in the equilibrium of part B due to the dependence of $\gamma_\mathrm{c}$ on $\rho_{44}$. In part C, where the atom cloud is again only irradiated by the weak probe laser, the process is very similar to what happens in part A with the exception that there is a small fraction of population is state $\ket{3}$. Consequently, the absorption level of the probe laser is smaller, because atoms are missing from the ground state.

In the case of initial population in the selected Rydberg state $\ket{3}$, the dynamics of the system only differ in parts A and B, provided B is long enough to reach steady state EIT. In part A, the absorption of the probe laser is close to zero, and slowly increases while a small fraction of the population decays from the Rydberg state $\ket{3}$. Since the lifetime of the Rydberg states is much longer than \SI{10}{\micro\second}, the amount of population transferred by spontaneous decay is small although not negligible on the \si{\micro\second} timescale of the pulse sequence. At the beginning of part B the population in the selected Rydberg state is transferred to the ground state $\ket{1}$ through a resonant transfer from state $\ket{3}$ to $\ket{2}$ induced by the coupling laser with Rabi frequency $\Omega_\mathrm{c}$ and the consecutive spontaneous emission from $\ket{2}$ to $\ket{1}$. For $\Omega_\mathrm{c}<\Gamma_{21}$, this process results in only a small increase in $\rho_{22}$, because a half Rabi cycle induced by $\Omega_\mathrm{c}$ between states $\ket{2}$ and $\ket{3}$ would take longer than the lifetime of state $\ket{2}$ (see Fig.~\ref{fig:popalpha}(b)). Hence, assuming the initial population is either in the ground state $\ket{1}$ or the selected Rydberg state $\ket{3}$, we can determine the fractions by monitoring the absorption of the probe laser in part A of the time evolution. If there is a way to ensure that all the population missing from the ground state $\ket{1}$ is in the selected Rydberg state $\ket{3}$, then this is indeed sufficient. However, if the probability that a fraction of the population is in another state (for example, the interaction scheme to be realized contains more than one Rydberg state), the absorption level in part A of the time evolution is not enough in itself to give information about the population of the selected Rydberg state $\ket{3}$. 

\begin{figure}
	\includegraphics{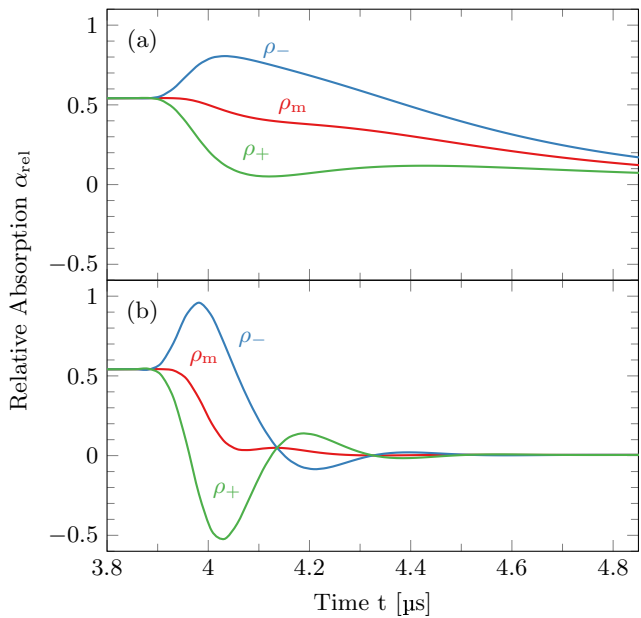}
	\caption{\label{fig:alpha_coherence}(Color online) Numerically calculated relative absorption $\alpha_{\mathrm{rel}}$ of the probe laser in the beginning of part B for atoms prepared in states $\rho_\mathrm{m}(t=0)=\frac{1}{2}(\ket{1}\bra{1}+\ket{3}\bra{3})$ and $\rho_\pm(t=0)=\frac{1}{4}(\ket{1}\pm\ket{3})(\bra{1}\pm\bra{3})$. The parameters used for the calculation are the same as for Fig.~\ref{fig:popalpha} with $\Omega_\mathrm{c}<\Gamma_{21}$ in (a) and $\Omega_\mathrm{c}\approx\Gamma_{21}$ in (b). This distinct signature of coherent states is expected to be experimentally observable and even more pronounced if $\Omega_\mathrm{c}$ is larger.}
\end{figure}

As illustrated in Fig.~\ref{fig:popalpha} (b) and (c), the absorption of the probe laser in part A is the same for different initial states of the atoms as long as the population in the ground state is the same. In contrast, the dynamics and the equilibrium become significantly different in part B. Since only the population in the selected Rydberg state is transferred back to the ground state by the coupling field, the absorption level in part C also changes with the initial population in $\ket{3}$. 

Another result of the simulations is the possibility to obtain information on the initial coherence of the system. The time evolution of the relative absorption in the beginning of part B for three different initial preparations of the atoms is shown in Fig.~\ref{fig:alpha_coherence}. These initial preparations consist of the same fraction of population in the ground state $\ket{1}$ and the selected Rydberg state $\ket{3}$, but the coherence between these two states is different. One of the initial preparations is the mixed state $\rho_\mathrm{m}(t=0)=\frac{1}{2}(\ket{1}\bra{1}+\ket{3}\bra{3})$, while the other two preparations are $\rho_\mathrm{\pm}(t=0)=\frac{1}{4}(\ket{1}\pm\ket{3})(\bra{1}\pm\bra{3})$. Comparing numerical calculations for these three cases, we find significant changes in the beginning of part B, where the EIT did not yet reach equilibrium. The absorption level during the rest of the pulse sequence is not sensitive to the initial coherence. If the Rabi frequency $\Omega_\mathrm{c}$ of the coupling laser is on the order of $\Gamma_{23}$ or higher, oscillations of the absorption signal can be observed.

\section{Experimental setup\label{sec:exp_setup}}

For demonstrating the detection of Rydberg population with time-resolved EIT, we have conducted an experiment on a cloud of \num{\approx2e7} $^{87}$Rb atoms at a temperature of \SI{\approx150}{\micro\kelvin}. In this experiment the atoms are trapped in a magneto-optical trap (MOT), loaded to a magnetic quadrupole trap and then released. The time-resolved measurements are started after \SI{3}{\milli\second} of time of flight, in order to ensure that all magnetic fields have fully decayed while the effects of atomic motion are still negligible. The measurements are performed within \SI{30}{\micro\second} (excitation pulse and probe sequence, cf. Fig.~\ref{fig:PulseSequence}). The density and optical density at the center of the cloud, measured by absorption imaging, are \SI{5e9}{\per\cubic\centi\meter} and \num{1.7}, respectively.

\begin{figure}
	\includegraphics{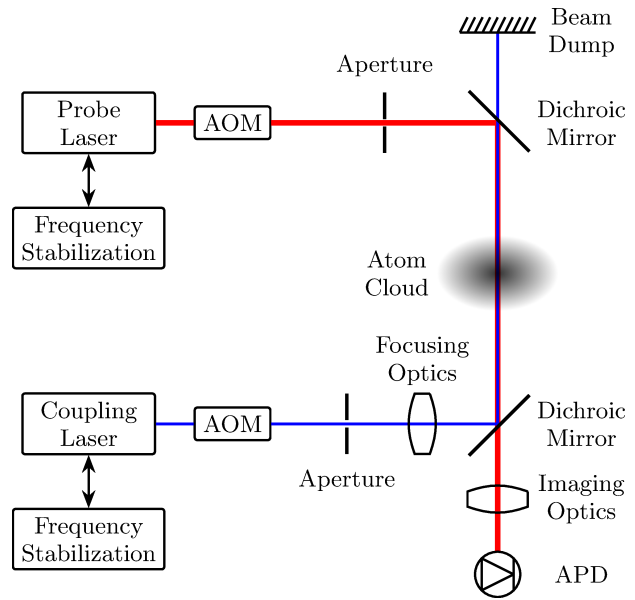}
	\caption{\label{fig:LaserSetup}(Color online) Experimental setup for time-resolved EIT measurements. The counter-propagating coupling (blue) and probe (red) beams are superimposed on the atom cloud and separated with dichroic mirrors. The transmission of the probe laser is detected by an avalanche photo diode (APD). The intensities of the lasers are controlled with acousto-optic modulators (AOM). Both laser frequencies are stabilized using a frequency comb.}
\end{figure}

The transitions from the ground state $\ket{5\mathrm{S}_{1/2},F=2}$ to the first excited state $\ket{5\mathrm{P}_{3/2},F=3}$ and from there to the selected Rydberg state $\ket{35\mathrm{S}_{1/2}}$ are driven by two lasers with wavelengths of \SI{\approx780}{\nano\meter} (red, probe) and \SI{\approx480}{\nano\meter} (blue, coupling), respectively (see Fig.~\ref{fig:ladder}). Additionally, we use a repumper to pump atoms from $\ket{5\mathrm{S}_{1/2},F=1}$ back to $\ket{5\mathrm{S}_{1/2},F=2}$ via $\ket{5\mathrm{P}_{3/2},F=2}$ during the whole pulse sequence. The frequencies of both lasers used in the experiment are referenced to a frequency comb and controlled with slow servo loops (\SI{<100}{\hertz} bandwidth).  The linewidths of both lasers are narrowed to less than \SI{2\pi\times20}{\kilo\hertz} with fast locks (\SI{>1}{\mega\hertz} bandwidth) to scanning Fabry-P\'{e}rot interferometers. As the Fabry-P\'{e}rot cavities are sensitive to acoustic noise, the effective linewidth for the experiment can be larger. The red and the blue lasers are aligned in a counter\-propa\-gating configuration (see Fig.~\ref{fig:LaserSetup}). We use an acousto-optic modulator (AOM) in each beam to create the intensity envelopes of the pulses. The switching time of the AOMs is \SI{50}{\nano\second} (\SI{20}{\percent} to \SI{80}{\percent} light intensity). The time-dependent intensity of the red laser (probe signal) is measured with an avalanche photo detector (Thorlabs APD120A/M) and recorded with a digital oscilloscope. The time resolution of the setup is \SI{20}{\nano\second}.

As shown in Fig.~\ref{fig:PulseSequence}, our measurement consists of three main parts. First we put all the atomic population into the Zeeman-sublevel of the ground state matching the polarization of the red laser by optically pumping for \SI{14.5}{\micro\second}. This pulse is also long enough that any transient effects due to the switching of the AOM wear off before the next pulse. Next, we prepare the initial state of the atomic cloud with an excitation pulse. We always apply the red laser for \SI{2}{\micro\second} with a Rabi frequency of \SI{2\pi\times3.3}{\mega\hertz}. If we want no population to be transferred to the Rydberg state, the blue laser remains switched off and due to the very short lifetime of the first excited state, practically the entire population remains in the ground state. Applying the blue laser with a Rabi frequency of \SI{2\pi\times1.8}{MHz} for up to \SI{2}{\micro\second} causes a fraction of the population to be transferred to the Rydberg state, with the most atoms being excited in the case of the full \SI{2}{\micro\second} pulse. Due to the high Rabi frequency of the red laser there is no coherent excitation of Rydberg atoms. In the third part (probe sequence in Fig.~\ref{fig:PulseSequence}), we use the red laser at a low intensity as the probe laser during time intervals A, B and C together with the blue laser as the coupling laser during time interval B (EIT pulse). 

As a reference, we always add one experimental cycle without atoms in order to measure the intensity $I_\mathrm{ref}(t)$ of the red laser with the photodetector and another one with atoms but no excitation pulse and no coupling pulse to normalize the data later on. The parameters are then varied from cycle to cycle and the transmitted intensity $I_\mathrm{T}(t)$ of the probe light after passing through the cloud is measured. The experiment is repeated several times for each set of parameters to reduce photo diode noise. The optical density OD is calculated as follows:
\begin{equation}
	\mathrm{OD}(t)=-\ln\left(\frac{I_\mathrm{T}(t)}{I_\mathrm{ref}(t)}\right).\label{eqn:OD}
\end{equation}
The resulting OD datasets are then normalized by dividing them by the OD dataset that had no excitation pulse and no EIT pulse [OD$_0(t)$]: 
\begin{equation}
	\mathrm{OD}_\mathrm{rel}(t) = \frac{\mathrm{OD}(t)}{\mathrm{OD}_0(t)} \mathrel{\hat{=}} \frac{\alpha(t)}{\alpha_0}.\label{eqn:ODrel}
\end{equation}
This relative optical density OD$_\mathrm{rel}$ is comparable to the relative absorption $\alpha(t)/\alpha_0$ that is calculated in the numerical simulation. To reduce the effect of the acoustic noise on our Fabry-P\'{e}rot cavities, we selected the 30 datasets where the mean transparency in the EIT pulse between \SI{7}{\micro\second} and \SI{9}{\micro\second} was maximal. We observe that in all measurements the relative optical density eventually returns to the level before the excitation pulse, ensuring that ionization effects are negligible. 

In general this scheme is applicable in situations where the optical density can be precisely measured. The resolution is limited by technical noise from the photodiode and the digital resolution of the subsequent data acquisition system. For low optical densities it is necessary to detect not only the absorption of the atoms, but to resolve the EIT signal as well. At high optical densities the constant resolution of the intensity measurement additionally leads to a lower resolution of the optical density due to the logarithmic scaling in Eq.~\eqref{eqn:OD}. For sufficient averaging, we estimate the presented scheme to be applicable in the range of optical densities between \num{\approx0.1} and \num{\approx4}. In principal, the optical density can be lowered by detuning the probe laser while maintaining the two photon resonance for the EIT condition, which on the other hand decreases the contrast for the EIT signal and therefore only allows for a limited extension of the range. 

\begin{figure}
	\includegraphics{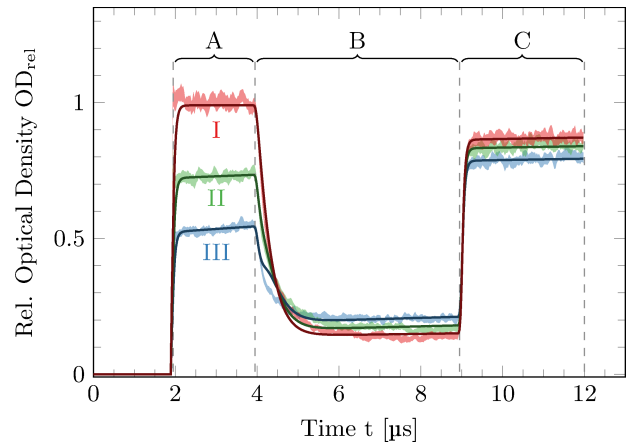}
	\caption{\label{fig:od_exp}(Color online) Measured optical density for the dynamics during the probe sequence. The shaded areas are \SI{95}{\percent} confidence intervals for the relative optical densities OD$_\mathrm{rel}(t)$ obtained from the measurements by applying Eqs.~\eqref{eqn:OD} and~\eqref{eqn:ODrel} for three durations of the excitation pulse (\SI{0}{\micro\second} for I, \SI{0.5}{\micro\second} for II and \SI{2}{\micro\second} for III). Solid lines represent fit results for the simulated relative absorption $\alpha(t)/\alpha_0$. The reversed order of the absorption signals in part B can be explained by dipole-dipole interactions.}
\end{figure}

\section{Detection of the initial population: experimental results\label{sec:expres}}

We demonstrate the optical detection of population for three different initial Rydberg excitation pulses. The experimental results for the optical density are shown in Fig.~\ref{fig:od_exp}. In order to compare these results to the model we calculate values for the decay rates matching the chosen combination of states in our experiment (see Fig.~\ref{fig:ladder}). We calculate the spontaneous emission rates using the wavefunctions calculated in \cite{Grimmel.2015}. $\Gamma_{32}$ is approximated by summing the spontaneous decay rates from $\ket{35\mathrm{S}_{1/2}}$ to all $n\mathrm{P}_j$ states, which results in 
\begin{equation}
	\Gamma_{32}=\sum\limits_{n>5}\Gamma_{\mathrm{sp},35\mathrm{S}\rightarrow n\mathrm{P}}=2\pi\times\SI{3.9}{\kilo\hertz}. 
\end{equation}
The main contribution comes from $\ket{5\mathrm{P}_{3/2}}$ ($\Gamma_{\mathrm{sp},35\mathrm{S}\rightarrow 5\mathrm{P}}=\SI{2\pi\times1.2}{\kilo\hertz}$) and other low-lying, fast-decaying states. The spontaneous decay rate $\Gamma_{34,\mathrm{sp}}$ is given by the rate $\Gamma_{\mathrm{sp},35\mathrm{S}\rightarrow 34\mathrm{P}}=2\pi\times\SI{16.8}{\hertz}$. Here we take only the strongest superradiant transition into account. The transition rate $\Gamma_{34,\mathrm{bb}}$ is approximated by a sum over all transition rates induced by blackbody radiation from $\ket{35\mathrm{S}_{1/2}}$ to all $n\mathrm{P}_j$ states
\begin{equation}
	\Gamma_{34,\mathrm{bb}}=\sum\limits_{n>5}\Gamma_{\mathrm{bb},35\mathrm{S}\rightarrow n\mathrm{P}}=2\pi\times\SI{2.7}{\kilo\hertz}, 
\end{equation}
in which the rates $\Gamma_{\mathrm{bb},35\mathrm{S}\rightarrow n\mathrm{P}}$ are calculated as in~\cite{Beterov.2009,Beterov.2009.Err}. The main contribution to $\Gamma_{34,\mathrm{bb}}$ comes from neighboring Ryd\-berg states. For the transitions induced by blackbody radiation a temperature of \SI{300}{\kelvin} is assumed. The preceding approximations for the decay rates ensure that the total decay out of the state $\ket{35\mathrm{S}_{1/2}}$ is modeled correctly. Similar to the calculation of $\Gamma_{32}$ we obtain 
\begin{equation}
	\Gamma_{41}=\sum\limits_{n>5}\Gamma_{\mathrm{sp},34\mathrm{P}\rightarrow n\mathrm{S}}=2\pi\times\SI{0.8}{\kilo\hertz}. 
\end{equation}
The Rabi frequencies of the two lasers $\Omega_\mathrm{p}=\SI{2\pi\times0.83}{\mega\hertz}$ and $\Omega_\mathrm{c}=\SI{2\pi\times2.10}{\mega\hertz}$ and the coherence decay $\gamma_\mathrm{c}=\SI{2\pi\times112}{\kilo\hertz}$ are fitted to the dataset without Rydberg excitation (see curve 'I' in Fig.~\ref{fig:od_exp}). The Rabi frequencies are consistent with estimates based on beam power and geometry. The noise $\gamma_\mathrm{c}$ is mainly caused by the acoustic noise on the Fabry-P\'{e}rot cavities. As the fit is only sensitive to $\gamma_\mathrm{p}+\gamma_\mathrm{c}$ and not to the single values, we choose $\gamma_\mathrm{p}=\SI{2\pi\times20}{\kilo\hertz}$.

\begin{figure}
	\includegraphics{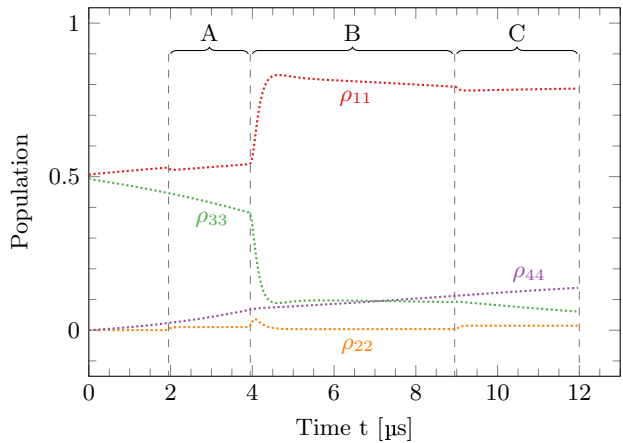}
	\caption{\label{fig:PopsExperiment}(Color online) Dynamics of the population of the four states ($\rho_{11}\hat{=}5\mathrm{S}_{1/2}$, $\rho_{22}\hat{=}5\mathrm{P}_{3/2}$, $\rho_{33}\hat{=}35\mathrm{S}_{1/2}$, $\rho_{44}\hat{=}\mathrm{Reservoir}$) retrieved from the fit to the dataset with high excitation (see Fig.~\ref{fig:od_exp}). The colors match those of the states in Fig.~\ref{fig:popalpha}. The population of the Rydberg state (35S$_{1/2}$) starts to decay immediately after excitation at $t=0$.} 
\end{figure}

An analysis of the datasets with the same conditions, but with Rydberg excitation, allows us to fit the fraction of atoms excited to a Rydberg state (\num{0.494+-0.008} and \num{0.284+-0.01}) within interval A, where only the red laser is on. The superradiance parameter $p_\mathrm{sup}=\num{7.9+-0.8e3}$ is fitted to parts B and C and the dipole-dipole interaction parameter $\gamma_{3,\mathrm{dd}}$ is adjusted in part B. The former scales with the absolute atom number while the latter scales with the atomic density. From the model one can now derive the time-resolved populations of the participating states as can be seen in Fig.~\ref{fig:PopsExperiment}. The accuracy of these populations is within \num{+-0.01} compared with values calculated using a variation method.

\section{Conclusion\label{sec:sum}}

We have demonstrated the all-optical detection of Rydberg population in a dilute gas, which is an alternative to the methods based on field ionization. Our results show that Rydberg population fractions can be measured with an accuracy of \num{0.01}. By comparing the dynamics of the measured optical densities to our numerical simulations we have quantified the decoherence effects occurring in the system, namely blackbody radiation induced transitions, superradiant decay and inhomogeneous broadening due to dipole-dipole interactions.

From our simulations we conclude that the detection scheme can also be used to obtain information on the coherence between the ground state and the Rydberg state. The numerical results predict that future studies using a coherent excitation method and experimental parameters similar to our experiment will be able to detect the initial and time-dependent coherence.

\appendix* 

\begin{acknowledgments}
The authors thank Daniel Cano and Florian Jessen for the design and setup of the experimental chamber and Claus Zimmermann for valuable discussions. This work was financially supported by the FET-Open Xtrack Project HAIRS and the Carl Zeiss Stiftung. N\'{o}ra S\'{a}ndor acknowledges financial support from the framework of T\'{A}MOP-4.2.4.A/2-11/1-2012-0001 'National Excellence Program'. 
\end{acknowledgments}


\begin{thebibliography}{26}%
\makeatletter
\providecommand \@ifxundefined [1]{%
 \@ifx{#1\undefined}
}%
\providecommand \@ifnum [1]{%
 \ifnum #1\expandafter \@firstoftwo
 \else \expandafter \@secondoftwo
 \fi
}%
\providecommand \@ifx [1]{%
 \ifx #1\expandafter \@firstoftwo
 \else \expandafter \@secondoftwo
 \fi
}%
\providecommand \natexlab [1]{#1}%
\providecommand \enquote  [1]{``#1''}%
\providecommand \bibnamefont  [1]{#1}%
\providecommand \bibfnamefont [1]{#1}%
\providecommand \citenamefont [1]{#1}%
\providecommand \href@noop [0]{\@secondoftwo}%
\providecommand \href [0]{\begingroup \@sanitize@url \@href}%
\providecommand \@href[1]{\@@startlink{#1}\@@href}%
\providecommand \@@href[1]{\endgroup#1\@@endlink}%
\providecommand \@sanitize@url [0]{\catcode `\\12\catcode `\$12\catcode
  `\&12\catcode `\#12\catcode `\^12\catcode `\_12\catcode `\%12\relax}%
\providecommand \@@startlink[1]{}%
\providecommand \@@endlink[0]{}%
\providecommand \url  [0]{\begingroup\@sanitize@url \@url }%
\providecommand \@url [1]{\endgroup\@href {#1}{\urlprefix }}%
\providecommand \urlprefix  [0]{URL }%
\providecommand \Eprint [0]{\href }%
\providecommand \doibase [0]{http://dx.doi.org/}%
\providecommand \selectlanguage [0]{\@gobble}%
\providecommand \bibinfo  [0]{\@secondoftwo}%
\providecommand \bibfield  [0]{\@secondoftwo}%
\providecommand \translation [1]{[#1]}%
\providecommand \BibitemOpen [0]{}%
\providecommand \bibitemStop [0]{}%
\providecommand \bibitemNoStop [0]{.\EOS\space}%
\providecommand \EOS [0]{\spacefactor3000\relax}%
\providecommand \BibitemShut  [1]{\csname bibitem#1\endcsname}%
\let\auto@bib@innerbib\@empty
\bibitem [{\citenamefont {Saffman}\ \emph {et~al.}(2010)\citenamefont
  {Saffman}, \citenamefont {Walker},\ and\ \citenamefont
  {M\o{}lmer}}]{Saffman.2010}%
  \BibitemOpen
  \bibfield  {author} {\bibinfo {author} {\bibfnamefont {M.}~\bibnamefont
  {Saffman}}, \bibinfo {author} {\bibfnamefont {T.~G.}\ \bibnamefont {Walker}},
  \ and\ \bibinfo {author} {\bibfnamefont {K.}~\bibnamefont {M\o{}lmer}},\
  }\href {\doibase 10.1103/RevModPhys.82.2313} {\bibfield  {journal} {\bibinfo
  {journal} {Rev. Mod. Phys.}\ }\textbf {\bibinfo {volume} {82}},\ \bibinfo
  {pages} {2313} (\bibinfo {year} {2010})}\BibitemShut {NoStop}%
\bibitem [{\citenamefont {Weimer}\ \emph {et~al.}(2010)\citenamefont {Weimer},
  \citenamefont {Muller}, \citenamefont {Lesanovsky}, \citenamefont {Zoller},\
  and\ \citenamefont {B\"{u}chler}}]{Weimer.2010}%
  \BibitemOpen
  \bibfield  {author} {\bibinfo {author} {\bibfnamefont {H.}~\bibnamefont
  {Weimer}}, \bibinfo {author} {\bibfnamefont {M.}~\bibnamefont {Muller}},
  \bibinfo {author} {\bibfnamefont {I.}~\bibnamefont {Lesanovsky}}, \bibinfo
  {author} {\bibfnamefont {P.}~\bibnamefont {Zoller}}, \ and\ \bibinfo {author}
  {\bibfnamefont {H.~P.}\ \bibnamefont {B\"{u}chler}},\ }\href {\doibase
  10.1038/nphys1614} {\bibfield  {journal} {\bibinfo  {journal} {Nat. Phys.}\
  }\textbf {\bibinfo {volume} {6}},\ \bibinfo {pages} {382} (\bibinfo {year}
  {2010})}\BibitemShut {NoStop}%
\bibitem [{\citenamefont {L{\"o}w}\ \emph {et~al.}(2012)\citenamefont
  {L{\"o}w}, \citenamefont {Weimer}, \citenamefont {Nipper}, \citenamefont
  {Balewski}, \citenamefont {Butscher}, \citenamefont {B{\"u}chler},\ and\
  \citenamefont {Pfau}}]{Low.2012}%
  \BibitemOpen
  \bibfield  {author} {\bibinfo {author} {\bibfnamefont {R.}~\bibnamefont
  {L{\"o}w}}, \bibinfo {author} {\bibfnamefont {H.}~\bibnamefont {Weimer}},
  \bibinfo {author} {\bibfnamefont {J.}~\bibnamefont {Nipper}}, \bibinfo
  {author} {\bibfnamefont {J.~B.}\ \bibnamefont {Balewski}}, \bibinfo {author}
  {\bibfnamefont {B.}~\bibnamefont {Butscher}}, \bibinfo {author}
  {\bibfnamefont {H.~P.}\ \bibnamefont {B{\"u}chler}}, \ and\ \bibinfo {author}
  {\bibfnamefont {T.}~\bibnamefont {Pfau}},\ }\href
  {http://stacks.iop.org/0953-4075/45/i=11/a=113001} {\bibfield  {journal}
  {\bibinfo  {journal} {J. Phys. B}\ }\textbf {\bibinfo {volume} {45}},\
  \bibinfo {pages} {113001} (\bibinfo {year} {2012})}\BibitemShut {NoStop}%
\bibitem [{\citenamefont {Tate}(2007)}]{Tate.2007}%
  \BibitemOpen
  \bibfield  {author} {\bibinfo {author} {\bibfnamefont {D.~A.}\ \bibnamefont
  {Tate}},\ }\href {\doibase 10.1103/PhysRevA.75.066502} {\bibfield  {journal}
  {\bibinfo  {journal} {Phys. Rev. A}\ }\textbf {\bibinfo {volume} {75}},\
  \bibinfo {pages} {066502} (\bibinfo {year} {2007})}\BibitemShut {NoStop}%
\bibitem [{\citenamefont {Caliri}\ and\ \citenamefont
  {Marcassa}(2007)}]{Caliri.2007}%
  \BibitemOpen
  \bibfield  {author} {\bibinfo {author} {\bibfnamefont {L.~L.}\ \bibnamefont
  {Caliri}}\ and\ \bibinfo {author} {\bibfnamefont {L.~G.}\ \bibnamefont
  {Marcassa}},\ }\href {\doibase 10.1103/PhysRevA.75.066503} {\bibfield
  {journal} {\bibinfo  {journal} {Phys. Rev. A}\ }\textbf {\bibinfo {volume}
  {75}},\ \bibinfo {pages} {066503} (\bibinfo {year} {2007})}\BibitemShut
  {NoStop}%
\bibitem [{\citenamefont {de~Oliveira}\ \emph {et~al.}(2002)\citenamefont
  {de~Oliveira}, \citenamefont {Mancini}, \citenamefont {Bagnato},\ and\
  \citenamefont {Marcassa}}]{Oliveira.2002}%
  \BibitemOpen
  \bibfield  {author} {\bibinfo {author} {\bibfnamefont {A.~L.}\ \bibnamefont
  {de~Oliveira}}, \bibinfo {author} {\bibfnamefont {M.~W.}\ \bibnamefont
  {Mancini}}, \bibinfo {author} {\bibfnamefont {V.~S.}\ \bibnamefont
  {Bagnato}}, \ and\ \bibinfo {author} {\bibfnamefont {L.~G.}\ \bibnamefont
  {Marcassa}},\ }\href {\doibase 10.1103/PhysRevA.65.031401} {\bibfield
  {journal} {\bibinfo  {journal} {Phys. Rev. A}\ }\textbf {\bibinfo {volume}
  {65}},\ \bibinfo {pages} {031401} (\bibinfo {year} {2002})}\BibitemShut
  {NoStop}%
\bibitem [{\citenamefont {Nascimento}\ \emph {et~al.}(2006)\citenamefont
  {Nascimento}, \citenamefont {Caliri}, \citenamefont {de~Oliveira},
  \citenamefont {Bagnato},\ and\ \citenamefont {Marcassa}}]{Nascimento.2006}%
  \BibitemOpen
  \bibfield  {author} {\bibinfo {author} {\bibfnamefont {V.~A.}\ \bibnamefont
  {Nascimento}}, \bibinfo {author} {\bibfnamefont {L.~L.}\ \bibnamefont
  {Caliri}}, \bibinfo {author} {\bibfnamefont {A.~L.}\ \bibnamefont
  {de~Oliveira}}, \bibinfo {author} {\bibfnamefont {V.~S.}\ \bibnamefont
  {Bagnato}}, \ and\ \bibinfo {author} {\bibfnamefont {L.~G.}\ \bibnamefont
  {Marcassa}},\ }\href {\doibase 10.1103/PhysRevA.74.054501} {\bibfield
  {journal} {\bibinfo  {journal} {Phys. Rev. A}\ }\textbf {\bibinfo {volume}
  {74}},\ \bibinfo {pages} {054501} (\bibinfo {year} {2006})}\BibitemShut
  {NoStop}%
\bibitem [{\citenamefont {Branden}\ \emph {et~al.}(2010)\citenamefont
  {Branden}, \citenamefont {Juhasz}, \citenamefont {Mahlokozera}, \citenamefont
  {Vesa}, \citenamefont {Wilson}, \citenamefont {Zheng}, \citenamefont
  {Kortyna},\ and\ \citenamefont {Tate}}]{Branden.2010}%
  \BibitemOpen
  \bibfield  {author} {\bibinfo {author} {\bibfnamefont {D.~B.}\ \bibnamefont
  {Branden}}, \bibinfo {author} {\bibfnamefont {T.}~\bibnamefont {Juhasz}},
  \bibinfo {author} {\bibfnamefont {T.}~\bibnamefont {Mahlokozera}}, \bibinfo
  {author} {\bibfnamefont {C.}~\bibnamefont {Vesa}}, \bibinfo {author}
  {\bibfnamefont {R.~O.}\ \bibnamefont {Wilson}}, \bibinfo {author}
  {\bibfnamefont {M.}~\bibnamefont {Zheng}}, \bibinfo {author} {\bibfnamefont
  {A.}~\bibnamefont {Kortyna}}, \ and\ \bibinfo {author} {\bibfnamefont
  {D.~A.}\ \bibnamefont {Tate}},\ }\href
  {http://stacks.iop.org/0953-4075/43/i=1/a=015002} {\bibfield  {journal}
  {\bibinfo  {journal} {J. Phys. B}\ }\textbf {\bibinfo {volume} {43}},\
  \bibinfo {pages} {015002} (\bibinfo {year} {2010})}\BibitemShut {NoStop}%
\bibitem [{\citenamefont {Cano}\ and\ \citenamefont
  {Fort\'agh}(2014)}]{Cano.2014}%
  \BibitemOpen
  \bibfield  {author} {\bibinfo {author} {\bibfnamefont {D.}~\bibnamefont
  {Cano}}\ and\ \bibinfo {author} {\bibfnamefont {J.}~\bibnamefont
  {Fort\'agh}},\ }\href {\doibase 10.1103/PhysRevA.89.043413} {\bibfield
  {journal} {\bibinfo  {journal} {Phys. Rev. A}\ }\textbf {\bibinfo {volume}
  {89}},\ \bibinfo {pages} {043413} (\bibinfo {year} {2014})}\BibitemShut
  {NoStop}%
\bibitem [{\citenamefont {Labuhn}\ \emph {et~al.}(2014)\citenamefont {Labuhn},
  \citenamefont {Ravets}, \citenamefont {Barredo}, \citenamefont {B\'eguin},
  \citenamefont {Nogrette}, \citenamefont {Lahaye},\ and\ \citenamefont
  {Browaeys}}]{Labuhn.2014}%
  \BibitemOpen
  \bibfield  {author} {\bibinfo {author} {\bibfnamefont {H.}~\bibnamefont
  {Labuhn}}, \bibinfo {author} {\bibfnamefont {S.}~\bibnamefont {Ravets}},
  \bibinfo {author} {\bibfnamefont {D.}~\bibnamefont {Barredo}}, \bibinfo
  {author} {\bibfnamefont {L.}~\bibnamefont {B\'eguin}}, \bibinfo {author}
  {\bibfnamefont {F.}~\bibnamefont {Nogrette}}, \bibinfo {author}
  {\bibfnamefont {T.}~\bibnamefont {Lahaye}}, \ and\ \bibinfo {author}
  {\bibfnamefont {A.}~\bibnamefont {Browaeys}},\ }\href {\doibase
  10.1103/PhysRevA.90.023415} {\bibfield  {journal} {\bibinfo  {journal} {Phys.
  Rev. A}\ }\textbf {\bibinfo {volume} {90}},\ \bibinfo {pages} {023415}
  (\bibinfo {year} {2014})}\BibitemShut {NoStop}%
\bibitem [{\citenamefont {Fleischhauer}\ \emph {et~al.}(2005)\citenamefont
  {Fleischhauer}, \citenamefont {Imamoglu},\ and\ \citenamefont
  {Marangos}}]{Fleischhauer.2005}%
  \BibitemOpen
  \bibfield  {author} {\bibinfo {author} {\bibfnamefont {M.}~\bibnamefont
  {Fleischhauer}}, \bibinfo {author} {\bibfnamefont {A.}~\bibnamefont
  {Imamoglu}}, \ and\ \bibinfo {author} {\bibfnamefont {J.~P.}\ \bibnamefont
  {Marangos}},\ }\href {\doibase 10.1103/RevModPhys.77.633} {\bibfield
  {journal} {\bibinfo  {journal} {Rev. Mod. Phys.}\ }\textbf {\bibinfo {volume}
  {77}},\ \bibinfo {pages} {633} (\bibinfo {year} {2005})}\BibitemShut
  {NoStop}%
\bibitem [{\citenamefont {Mohapatra}\ \emph {et~al.}(2007)\citenamefont
  {Mohapatra}, \citenamefont {Jackson},\ and\ \citenamefont
  {Adams}}]{Mohapatra.2007}%
  \BibitemOpen
  \bibfield  {author} {\bibinfo {author} {\bibfnamefont {A.~K.}\ \bibnamefont
  {Mohapatra}}, \bibinfo {author} {\bibfnamefont {T.~R.}\ \bibnamefont
  {Jackson}}, \ and\ \bibinfo {author} {\bibfnamefont {C.~S.}\ \bibnamefont
  {Adams}},\ }\href {\doibase 10.1103/PhysRevLett.98.113003} {\bibfield
  {journal} {\bibinfo  {journal} {Phys. Rev. Lett.}\ }\textbf {\bibinfo
  {volume} {98}},\ \bibinfo {pages} {113003} (\bibinfo {year}
  {2007})}\BibitemShut {NoStop}%
\bibitem [{\citenamefont {Mack}\ \emph {et~al.}(2011)\citenamefont {Mack},
  \citenamefont {Karlewski}, \citenamefont {Hattermann}, \citenamefont
  {H\"ockh}, \citenamefont {Jessen}, \citenamefont {Cano},\ and\ \citenamefont
  {Fort\'agh}}]{Mack.2011}%
  \BibitemOpen
  \bibfield  {author} {\bibinfo {author} {\bibfnamefont {M.}~\bibnamefont
  {Mack}}, \bibinfo {author} {\bibfnamefont {F.}~\bibnamefont {Karlewski}},
  \bibinfo {author} {\bibfnamefont {H.}~\bibnamefont {Hattermann}}, \bibinfo
  {author} {\bibfnamefont {S.}~\bibnamefont {H\"ockh}}, \bibinfo {author}
  {\bibfnamefont {F.}~\bibnamefont {Jessen}}, \bibinfo {author} {\bibfnamefont
  {D.}~\bibnamefont {Cano}}, \ and\ \bibinfo {author} {\bibfnamefont
  {J.}~\bibnamefont {Fort\'agh}},\ }\href {\doibase 10.1103/PhysRevA.83.052515}
  {\bibfield  {journal} {\bibinfo  {journal} {Phys. Rev. A}\ }\textbf {\bibinfo
  {volume} {83}},\ \bibinfo {pages} {052515} (\bibinfo {year}
  {2011})}\BibitemShut {NoStop}%
\bibitem [{\citenamefont {Weatherill}\ \emph {et~al.}(2008)\citenamefont
  {Weatherill}, \citenamefont {Pritchard}, \citenamefont {Abel}, \citenamefont
  {Bason}, \citenamefont {Mohapatra},\ and\ \citenamefont
  {Adams}}]{Weatherill.2008}%
  \BibitemOpen
  \bibfield  {author} {\bibinfo {author} {\bibfnamefont {K.~J.}\ \bibnamefont
  {Weatherill}}, \bibinfo {author} {\bibfnamefont {J.~D.}\ \bibnamefont
  {Pritchard}}, \bibinfo {author} {\bibfnamefont {R.~P.}\ \bibnamefont {Abel}},
  \bibinfo {author} {\bibfnamefont {M.~G.}\ \bibnamefont {Bason}}, \bibinfo
  {author} {\bibfnamefont {A.~K.}\ \bibnamefont {Mohapatra}}, \ and\ \bibinfo
  {author} {\bibfnamefont {C.~S.}\ \bibnamefont {Adams}},\ }\href
  {http://stacks.iop.org/0953-4075/41/i=20/a=201002} {\bibfield  {journal}
  {\bibinfo  {journal} {J. Phys. B}\ }\textbf {\bibinfo {volume} {41}},\
  \bibinfo {pages} {201002} (\bibinfo {year} {2008})}\BibitemShut {NoStop}%
\bibitem [{\citenamefont {Tauschinsky}\ \emph {et~al.}(2010)\citenamefont
  {Tauschinsky}, \citenamefont {Thijssen}, \citenamefont {Whitlock},
  \citenamefont {van Linden van~den Heuvell},\ and\ \citenamefont
  {Spreeuw}}]{Tauschinsky.2010}%
  \BibitemOpen
  \bibfield  {author} {\bibinfo {author} {\bibfnamefont {A.}~\bibnamefont
  {Tauschinsky}}, \bibinfo {author} {\bibfnamefont {R.~M.~T.}\ \bibnamefont
  {Thijssen}}, \bibinfo {author} {\bibfnamefont {S.}~\bibnamefont {Whitlock}},
  \bibinfo {author} {\bibfnamefont {H.~B.}\ \bibnamefont {van Linden van~den
  Heuvell}}, \ and\ \bibinfo {author} {\bibfnamefont {R.~J.~C.}\ \bibnamefont
  {Spreeuw}},\ }\href {\doibase 10.1103/PhysRevA.81.063411} {\bibfield
  {journal} {\bibinfo  {journal} {Phys. Rev. A}\ }\textbf {\bibinfo {volume}
  {81}},\ \bibinfo {pages} {063411} (\bibinfo {year} {2010})}\BibitemShut
  {NoStop}%
\bibitem [{\citenamefont {Bason}\ \emph {et~al.}(2010)\citenamefont {Bason},
  \citenamefont {Tanasittikosol}, \citenamefont {Sargsyan}, \citenamefont
  {Mohapatra}, \citenamefont {Sarkisyan}, \citenamefont {Potvliege},\ and\
  \citenamefont {Adams}}]{Bason.2010}%
  \BibitemOpen
  \bibfield  {author} {\bibinfo {author} {\bibfnamefont {M.~G.}\ \bibnamefont
  {Bason}}, \bibinfo {author} {\bibfnamefont {M.}~\bibnamefont
  {Tanasittikosol}}, \bibinfo {author} {\bibfnamefont {A.}~\bibnamefont
  {Sargsyan}}, \bibinfo {author} {\bibfnamefont {A.~K.}\ \bibnamefont
  {Mohapatra}}, \bibinfo {author} {\bibfnamefont {D.}~\bibnamefont
  {Sarkisyan}}, \bibinfo {author} {\bibfnamefont {R.~M.}\ \bibnamefont
  {Potvliege}}, \ and\ \bibinfo {author} {\bibfnamefont {C.~S.}\ \bibnamefont
  {Adams}},\ }\href {http://stacks.iop.org/1367-2630/12/i=6/a=065015}
  {\bibfield  {journal} {\bibinfo  {journal} {New Journal of Physics}\ }\textbf
  {\bibinfo {volume} {12}},\ \bibinfo {pages} {065015} (\bibinfo {year}
  {2010})}\BibitemShut {NoStop}%
\bibitem [{\citenamefont {Hattermann}\ \emph {et~al.}(2012)\citenamefont
  {Hattermann}, \citenamefont {Mack}, \citenamefont {Karlewski}, \citenamefont
  {Jessen}, \citenamefont {Cano},\ and\ \citenamefont
  {Fort\'agh}}]{Hattermann.2012}%
  \BibitemOpen
  \bibfield  {author} {\bibinfo {author} {\bibfnamefont {H.}~\bibnamefont
  {Hattermann}}, \bibinfo {author} {\bibfnamefont {M.}~\bibnamefont {Mack}},
  \bibinfo {author} {\bibfnamefont {F.}~\bibnamefont {Karlewski}}, \bibinfo
  {author} {\bibfnamefont {F.}~\bibnamefont {Jessen}}, \bibinfo {author}
  {\bibfnamefont {D.}~\bibnamefont {Cano}}, \ and\ \bibinfo {author}
  {\bibfnamefont {J.}~\bibnamefont {Fort\'agh}},\ }\href {\doibase
  10.1103/PhysRevA.86.022511} {\bibfield  {journal} {\bibinfo  {journal} {Phys.
  Rev. A}\ }\textbf {\bibinfo {volume} {86}},\ \bibinfo {pages} {022511}
  (\bibinfo {year} {2012})}\BibitemShut {NoStop}%
\bibitem [{\citenamefont {Tauschinsky}\ \emph {et~al.}(2013)\citenamefont
  {Tauschinsky}, \citenamefont {Newell}, \citenamefont {van Linden van~den
  Heuvell},\ and\ \citenamefont {Spreeuw}}]{Tauschinsky.2013}%
  \BibitemOpen
  \bibfield  {author} {\bibinfo {author} {\bibfnamefont {A.}~\bibnamefont
  {Tauschinsky}}, \bibinfo {author} {\bibfnamefont {R.}~\bibnamefont {Newell}},
  \bibinfo {author} {\bibfnamefont {H.~B.}\ \bibnamefont {van Linden van~den
  Heuvell}}, \ and\ \bibinfo {author} {\bibfnamefont {R.~J.~C.}\ \bibnamefont
  {Spreeuw}},\ }\href {\doibase 10.1103/PhysRevA.87.042522} {\bibfield
  {journal} {\bibinfo  {journal} {Phys. Rev. A}\ }\textbf {\bibinfo {volume}
  {87}},\ \bibinfo {pages} {042522} (\bibinfo {year} {2013})}\BibitemShut
  {NoStop}%
\bibitem [{\citenamefont {Grimmel}\ \emph {et~al.}(2015)\citenamefont
  {Grimmel}, \citenamefont {Mack}, \citenamefont {Karlewski}, \citenamefont
  {Jessen}, \citenamefont {Reinschmidt}, \citenamefont {S\'{a}ndor},\ and\
  \citenamefont {Fort\'agh}}]{Grimmel.2015}%
  \BibitemOpen
  \bibfield  {author} {\bibinfo {author} {\bibfnamefont {J.}~\bibnamefont
  {Grimmel}}, \bibinfo {author} {\bibfnamefont {M.}~\bibnamefont {Mack}},
  \bibinfo {author} {\bibfnamefont {F.}~\bibnamefont {Karlewski}}, \bibinfo
  {author} {\bibfnamefont {F.}~\bibnamefont {Jessen}}, \bibinfo {author}
  {\bibfnamefont {M.}~\bibnamefont {Reinschmidt}}, \bibinfo {author}
  {\bibfnamefont {N.}~\bibnamefont {S\'{a}ndor}}, \ and\ \bibinfo {author}
  {\bibfnamefont {J.}~\bibnamefont {Fort\'agh}},\ }\href@noop {} {\bibfield
  {journal} {\bibinfo  {journal} {To be published}\ } (\bibinfo {year}
  {2015})}\BibitemShut {NoStop}%
\bibitem [{\citenamefont {G\"unter}\ \emph {et~al.}(2012)\citenamefont
  {G\"unter}, \citenamefont {{Robert-de-Saint-Vincent}}, \citenamefont
  {Schempp}, \citenamefont {Hofmann}, \citenamefont {Whitlock},\ and\
  \citenamefont {Weidem\"uller}}]{Gunter.2012}%
  \BibitemOpen
  \bibfield  {author} {\bibinfo {author} {\bibfnamefont {G.}~\bibnamefont
  {G\"unter}}, \bibinfo {author} {\bibfnamefont {M.}~\bibnamefont
  {{Robert-de-Saint-Vincent}}}, \bibinfo {author} {\bibfnamefont
  {H.}~\bibnamefont {Schempp}}, \bibinfo {author} {\bibfnamefont {C.~S.}\
  \bibnamefont {Hofmann}}, \bibinfo {author} {\bibfnamefont {S.}~\bibnamefont
  {Whitlock}}, \ and\ \bibinfo {author} {\bibfnamefont {M.}~\bibnamefont
  {Weidem\"uller}},\ }\href {\doibase 10.1103/PhysRevLett.108.013002}
  {\bibfield  {journal} {\bibinfo  {journal} {Phys. Rev. Lett.}\ }\textbf
  {\bibinfo {volume} {108}},\ \bibinfo {pages} {013002} (\bibinfo {year}
  {2012})}\BibitemShut {NoStop}%
\bibitem [{\citenamefont {G{\"u}nter}\ \emph {et~al.}(2013)\citenamefont
  {G{\"u}nter}, \citenamefont {Schempp}, \citenamefont
  {{Robert-de-Saint-Vincent}}, \citenamefont {Gavryusev}, \citenamefont
  {Helmrich}, \citenamefont {Hofmann}, \citenamefont {Whitlock},\ and\
  \citenamefont {Weidem{\"u}ller}}]{Gunter.2013}%
  \BibitemOpen
  \bibfield  {author} {\bibinfo {author} {\bibfnamefont {G.}~\bibnamefont
  {G{\"u}nter}}, \bibinfo {author} {\bibfnamefont {H.}~\bibnamefont {Schempp}},
  \bibinfo {author} {\bibfnamefont {M.}~\bibnamefont
  {{Robert-de-Saint-Vincent}}}, \bibinfo {author} {\bibfnamefont
  {V.}~\bibnamefont {Gavryusev}}, \bibinfo {author} {\bibfnamefont
  {S.}~\bibnamefont {Helmrich}}, \bibinfo {author} {\bibfnamefont {C.~S.}\
  \bibnamefont {Hofmann}}, \bibinfo {author} {\bibfnamefont {S.}~\bibnamefont
  {Whitlock}}, \ and\ \bibinfo {author} {\bibfnamefont {M.}~\bibnamefont
  {Weidem{\"u}ller}},\ }\href {\doibase 10.1126/science.1244843} {\bibfield
  {journal} {\bibinfo  {journal} {Science}\ }\textbf {\bibinfo {volume}
  {342}},\ \bibinfo {pages} {954} (\bibinfo {year} {2013})}\BibitemShut
  {NoStop}%
\bibitem [{\citenamefont {Beterov}\ \emph
  {et~al.}(2009{\natexlab{a}})\citenamefont {Beterov}, \citenamefont
  {Ryabtsev}, \citenamefont {Tretyakov},\ and\ \citenamefont
  {Entin}}]{Beterov.2009}%
  \BibitemOpen
  \bibfield  {author} {\bibinfo {author} {\bibfnamefont {I.~I.}\ \bibnamefont
  {Beterov}}, \bibinfo {author} {\bibfnamefont {I.~I.}\ \bibnamefont
  {Ryabtsev}}, \bibinfo {author} {\bibfnamefont {D.~B.}\ \bibnamefont
  {Tretyakov}}, \ and\ \bibinfo {author} {\bibfnamefont {V.~M.}\ \bibnamefont
  {Entin}},\ }\href {\doibase 10.1103/PhysRevA.79.052504} {\bibfield  {journal}
  {\bibinfo  {journal} {Phys. Rev. A}\ }\textbf {\bibinfo {volume} {79}},\
  \bibinfo {pages} {052504} (\bibinfo {year} {2009}{\natexlab{a}})}\BibitemShut
  {NoStop}%
\bibitem [{\citenamefont {Beterov}\ \emph
  {et~al.}(2009{\natexlab{b}})\citenamefont {Beterov}, \citenamefont
  {Ryabtsev}, \citenamefont {Tretyakov},\ and\ \citenamefont
  {Entin}}]{Beterov.2009.Err}%
  \BibitemOpen
  \bibfield  {author} {\bibinfo {author} {\bibfnamefont {I.~I.}\ \bibnamefont
  {Beterov}}, \bibinfo {author} {\bibfnamefont {I.~I.}\ \bibnamefont
  {Ryabtsev}}, \bibinfo {author} {\bibfnamefont {D.~B.}\ \bibnamefont
  {Tretyakov}}, \ and\ \bibinfo {author} {\bibfnamefont {V.~M.}\ \bibnamefont
  {Entin}},\ }\href {\doibase 10.1103/PhysRevA.80.059902} {\bibfield  {journal}
  {\bibinfo  {journal} {Phys. Rev. A}\ }\textbf {\bibinfo {volume} {80}},\
  \bibinfo {pages} {059902(E)} (\bibinfo {year}
  {2009}{\natexlab{b}})}\BibitemShut {NoStop}%
\bibitem [{\citenamefont {Wang}\ \emph {et~al.}(2007)\citenamefont {Wang},
  \citenamefont {Yelin}, \citenamefont {C\^ot\'e}, \citenamefont {Eyler},
  \citenamefont {Farooqi}, \citenamefont {Gould}, \citenamefont
  {Ko\ifmmode~\check{s}\else \v{s}\fi{}trun}, \citenamefont {Tong},\ and\
  \citenamefont {Vrinceanu}}]{Wang.2007}%
  \BibitemOpen
  \bibfield  {author} {\bibinfo {author} {\bibfnamefont {T.}~\bibnamefont
  {Wang}}, \bibinfo {author} {\bibfnamefont {S.~F.}\ \bibnamefont {Yelin}},
  \bibinfo {author} {\bibfnamefont {R.}~\bibnamefont {C\^ot\'e}}, \bibinfo
  {author} {\bibfnamefont {E.~E.}\ \bibnamefont {Eyler}}, \bibinfo {author}
  {\bibfnamefont {S.~M.}\ \bibnamefont {Farooqi}}, \bibinfo {author}
  {\bibfnamefont {P.~L.}\ \bibnamefont {Gould}}, \bibinfo {author}
  {\bibfnamefont {M.}~\bibnamefont {Ko\ifmmode~\check{s}\else \v{s}\fi{}trun}},
  \bibinfo {author} {\bibfnamefont {D.}~\bibnamefont {Tong}}, \ and\ \bibinfo
  {author} {\bibfnamefont {D.}~\bibnamefont {Vrinceanu}},\ }\href {\doibase
  10.1103/PhysRevA.75.033802} {\bibfield  {journal} {\bibinfo  {journal} {Phys.
  Rev. A}\ }\textbf {\bibinfo {volume} {75}},\ \bibinfo {pages} {033802}
  (\bibinfo {year} {2007})}\BibitemShut {NoStop}%
\bibitem [{\citenamefont {Anderson}\ \emph {et~al.}(2002)\citenamefont
  {Anderson}, \citenamefont {Robinson}, \citenamefont {Martin},\ and\
  \citenamefont {Gallagher}}]{Anderson.2002}%
  \BibitemOpen
  \bibfield  {author} {\bibinfo {author} {\bibfnamefont {W.~R.}\ \bibnamefont
  {Anderson}}, \bibinfo {author} {\bibfnamefont {M.~P.}\ \bibnamefont
  {Robinson}}, \bibinfo {author} {\bibfnamefont {J.~D.~D.}\ \bibnamefont
  {Martin}}, \ and\ \bibinfo {author} {\bibfnamefont {T.~F.}\ \bibnamefont
  {Gallagher}},\ }\href {\doibase 10.1103/PhysRevA.65.063404} {\bibfield
  {journal} {\bibinfo  {journal} {Phys. Rev. A}\ }\textbf {\bibinfo {volume}
  {65}},\ \bibinfo {pages} {063404} (\bibinfo {year} {2002})}\BibitemShut
  {NoStop}%
\bibitem [{\citenamefont {Day}\ \emph {et~al.}(2008)\citenamefont {Day},
  \citenamefont {Brekke},\ and\ \citenamefont {Walker}}]{Day.2008}%
  \BibitemOpen
  \bibfield  {author} {\bibinfo {author} {\bibfnamefont {J.~O.}\ \bibnamefont
  {Day}}, \bibinfo {author} {\bibfnamefont {E.}~\bibnamefont {Brekke}}, \ and\
  \bibinfo {author} {\bibfnamefont {T.~G.}\ \bibnamefont {Walker}},\ }\href
  {\doibase 10.1103/PhysRevA.77.052712} {\bibfield  {journal} {\bibinfo
  {journal} {Phys. Rev. A}\ }\textbf {\bibinfo {volume} {77}},\ \bibinfo
  {pages} {052712} (\bibinfo {year} {2008})}\BibitemShut {NoStop}%
\end{thebibliography}
\end{document}